\documentclass[aps,prl,twocolumn,letterpaper,superscriptaddress]{revtex4} %preprint
\usepackage{dsfont}
\usepackage{amsmath}
\usepackage{graphicx}
\usepackage{multirow}
\usepackage{color}
\usepackage{colortbl}
\usepackage{xcolor}  % for writting parts in color, e.g. in red

\usepackage[colorlinks=true,citecolor=black,urlcolor=black,linkcolor=black]{hyperref} % choice of color for hyperref links

\usepackage[FIGTOPCAP, hang, raggedright, nooneline]{subfigure}
\usepackage{extarrows}
\usepackage{ulem}

\makeindex

\begin{document}

\title{Ultracold atoms out of equilibrium}
%\markboth{Langen, Geiger \& Schmiedmayer}{Ultracold atoms out of equilibrium}

\author{T. Langen} 
\email{tlangen@ati.ac.at}
\affiliation{Vienna Center for Quantum Science and Technology, Atominstitut, TU Wien, Stadionallee 2, 1020 Vienna, Austria}

\author{R. Geiger}
\affiliation{Vienna Center for Quantum Science and Technology, Atominstitut, TU Wien, Stadionallee 2, 1020 Vienna, Austria}

\author{J. Schmiedmayer} 
\email{schmiedmayer@atomchip.org}
\affiliation{Vienna Center for Quantum Science and Technology, Atominstitut, TU Wien, Stadionallee 2, 1020 Vienna, Austria}

%\begin{keywords}
%Ultracold Quantum Gases, Non-Equilibrium Dynamics, Quantum Many-Body Physics, Statistical Physics
%\end{keywords}

\begin{abstract}
The relaxation of isolated quantum many-body systems is a major unsolved problem connecting statistical and quantum physics. Studying such relaxation processes remains a challenge despite considerable efforts. Experimentally, it requires the creation and manipulation of well-controlled and truly isolated quantum systems. In this context, ultracold neutral atoms provide unique opportunities to understand non-equilibrium phenomena because of the large set of available methods to isolate, manipulate and probe these systems. Here, we give an overview of the rapid experimental progress that has been made in the field over the last years and highlight some of the questions which may be explored in the future. 
\end{abstract}

\maketitle

\section{Introduction}
Statistical mechanics provides a powerful description of the thermal properties of many-body systems. Typically, in this description the system under study is being coupled to a large reservoir, with which it can exchange particles or energy to reach a state of thermal equilibrium~\cite{HuangBook}. Moreover, it is implicitly assumed that this reservoir itself is in thermal equilibrium. If we do not want to invoke an even larger reservoir that thermalized the first one, we naturally arrive at the fundamental question if already a single, isolated many-body system can evolve in such a way that it reaches an (apparent) thermal equilibrium state. While this question is well understood for classical systems, the quantum case still lacks a general  description~\cite{Polkovnikov11}. 

The understanding of such non-equilibrium dynamics is not only the topic of intense research in quantum statistical physics, but also an open problem in diverse fields such as cosmology~\cite{Podolsky06,Kofman94,Arrizabalaga05}, high-energy physics~\cite{Berges04,BraunMunzinger01,Heinz02}, quantum information and condensed matter~\cite{Kollath07,Eckstein09,Moeckel10,Barnett10}, spanning virtually all energy, time and length scales. As a consequence, the term non-equilibrium dynamics encompasses many different protocols and phenomena. Topics that have been investigated theoretically range from dynamical phase transitions and the emergence of a thermodynamical description to transport phenomena and the interplay between dynamics and disorder~\cite{Barmettler09,Heyl13,Eckstein09,Muth10,Srednicki94,Rigol08,Reimann08,Popescu06}. 

A key challenge is the scarcity of experimental platforms to probe such relaxation dynamics in detail. These platforms have to be at the same time sufficiently well isolated from the environment and still accessible for experimental study. Over the last years, ultracold atomic gases have emerged as versatile model systems, as they combine these two essential prerequisites in a very unique fashion~\cite{PethickSmith,Lamacraft12}. In this review, we present a (non-exhaustive) overview of recent experiments that demonstrate how ultracold atoms can provide comprehensive insights into many aspects of non-equilibrium quantum many-body physics.

\section{Non-equilibrium dynamics of isolated quantum systems}

\begin{figure*}[htp]
	\centering
		\includegraphics[width=0.69\textwidth]{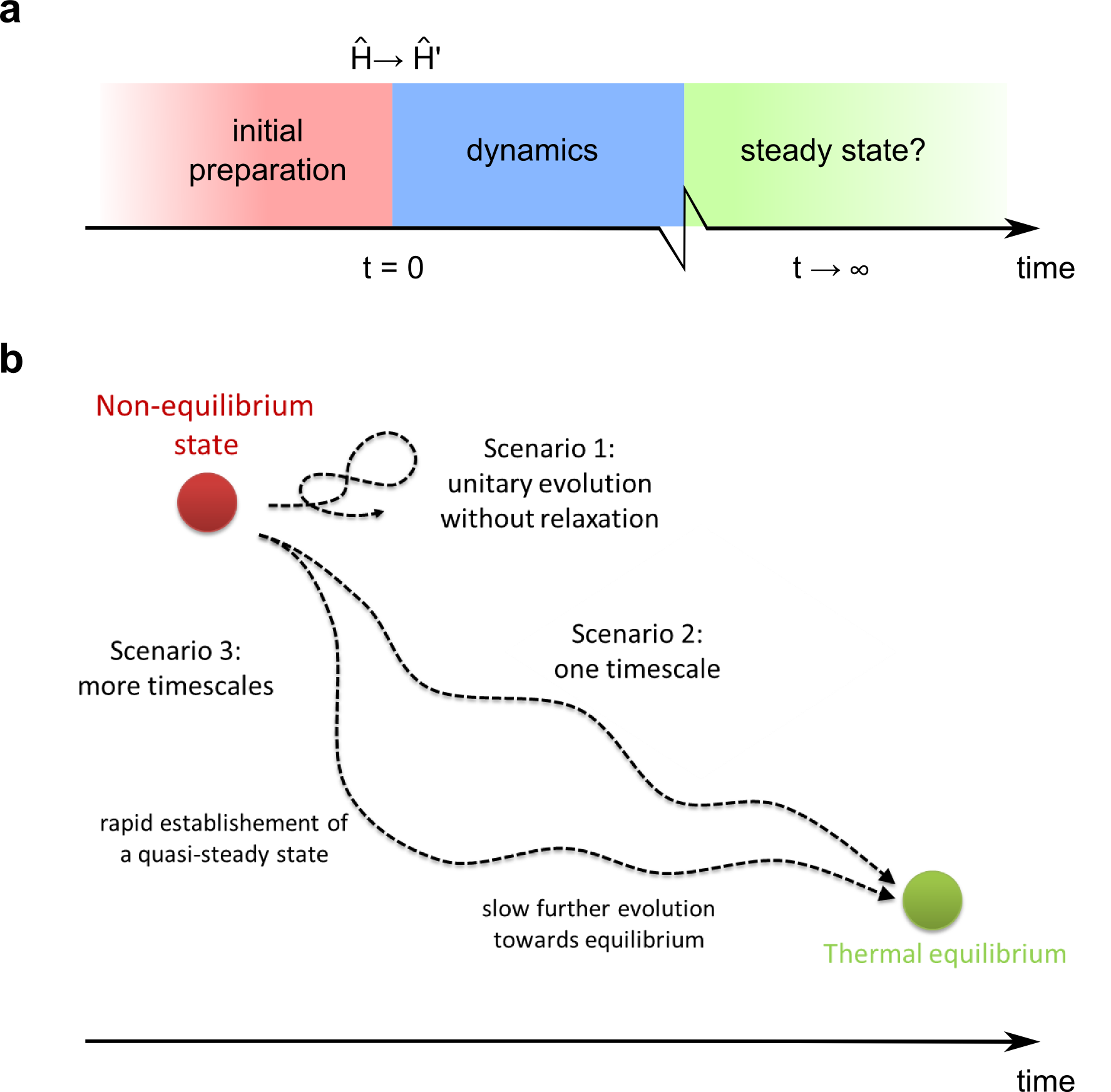}
	\caption{\textbf{(a)} The generic protocol of a quench. After the preparation of an initial state, the Hamiltonian of the system is rapidly changed, resulting in a non-equilibrium state. This induces a dynamical evolution. It is an open question whether a steady state emerges and if so, what the properties of this state are. \textbf{(b)} Several scenarios are conceivable for the dynamics. Following the strictly unitary evolution of quantum mechanics, thermal equilibrium can never be reached. However, alternative scenarios have been put forward, where the expectation values of observables might come arbitrarily close to their thermal values. The corresponding relaxation could happen on a single timescale, but also be more complex with one or more intermediate states that already share certain properties with the thermal equilibrium state. Figure adapted from~\cite{LangenPhD}.
	}
	\label{fig1}
\end{figure*}

We start with a brief outline of the generic theoretical expectations for relaxation processes in quantum many-body systems. A particularly useful protocol is the quantum quench, where the dynamical response of a system to a sudden perturbation is studied (Fig.~\ref{fig1}). Imagine a rapid change of a generic quantum many-body system's Hamiltonian $H\rightarrow H'$ at some time $t_0$. Then, the expectation value of an observable $O$ after the quench is given by $\langle O(t)\rangle=\langle \psi(t)|O|\psi(t)\rangle$. The many-body wave function $|\psi(t)\rangle$ evolves in time as $|\psi(t)\rangle = \exp(-iH't/\hbar)|\psi_0\rangle$ under the influence of the new Hamiltonian $H'$. Except for the artificial case where the state $|\psi_0\rangle$ after the quench is an eigenstate of $H'$, this simple equation describes an extremely complex time evolution, for which no general solution exists. Consequently, it has so far only been possible to understand if and how isolated quantum systems relax towards equilibrium states for a very limited number of special cases~\cite{Polkovnikov11}. Thermalization would require a complete loss of all memory of the initial state $|\psi_0\rangle$. However, because the time-evolution in quantum mechanics is unitary, this loss of memory seems impossible. 

The apparent absence of thermalization is in stark contradiction to our common knowledge, which tells us that many quantum mechanical systems, including ultracold atoms, can be described by a thermal state. This paradox was noticed already shortly after the introduction of quantum mechanics, with the earliest attempts to resolve it dating back to the 1930s~\cite{Neumann29,Polkovnikov11}. The general key to resolving the paradox is the fact that the central role for our observations is not played by the many-body wave functions, but instead by the outcomes of the measurement process. In a single run, a large quantum many-body system will evolve in a unitary way, but when observables are measured, their expectation values might become arbitrarily close to the prediction of a thermal state. A particular well-studied scenario to grasp this intuition more formally is the eigenstate thermalization hypothesis (ETH)~\cite{Srednicki94,Deutsch91,Rigol08}, which conjectures that the initial state of a non-equilibrium evolution already has thermal properties on the level of individual many-body eigenstates. While an experimental observation is still lacking, numerical results in some systems (in particular ones with a chaotic classical limit) indicate that the ETH is fulfilled for generic few-body observables, i.e. observables that only act on a small subsystem of the total system~\cite{Rigol08,Rigol12}. An intuitive picture for this observation is that the isolated total system acts as an environment that thermalizes its few-body subsystems~\cite{Cramer08}. However, not all systems are expected to thermalize, the most notable exceptions being integrable~\cite{Rigol09} and localized systems~\cite{Nandkishore14}.

If we assume that some isolated quantum systems can appear (for all practical purposes) thermalized, the next important question is how the thermalization proceeds. For example, there might be partial relaxation only, where instead of a complete loss of memory of the initial state, the system only partially forgets the initial conditions. Also, there might be different stages of relaxation connected to different time scales. Such relaxation with different time scales has been predicted to occur in many systems~\cite{Podolsky06,Berges04,Kollath07,Barnett10,Arrizabalaga05,Moeckel10,Kitagawa11,Kollar11,Berges2007}, providing a striking example of the complexity of non-equilibrium dynamics in the quantum world.

\section{Studying non-equilibrium dynamics with cold atoms}

Atomic gases provide unique opportunities for studying these non-equilibrium processes in the laboratory. The quantum evolution can be observed on experimentally accessible timescales and the tunability of many parameters allow the realization of a multitude of different physical situations. 

\begin{figure*}[tb]
	\centering
		\includegraphics[width=.7\textwidth]{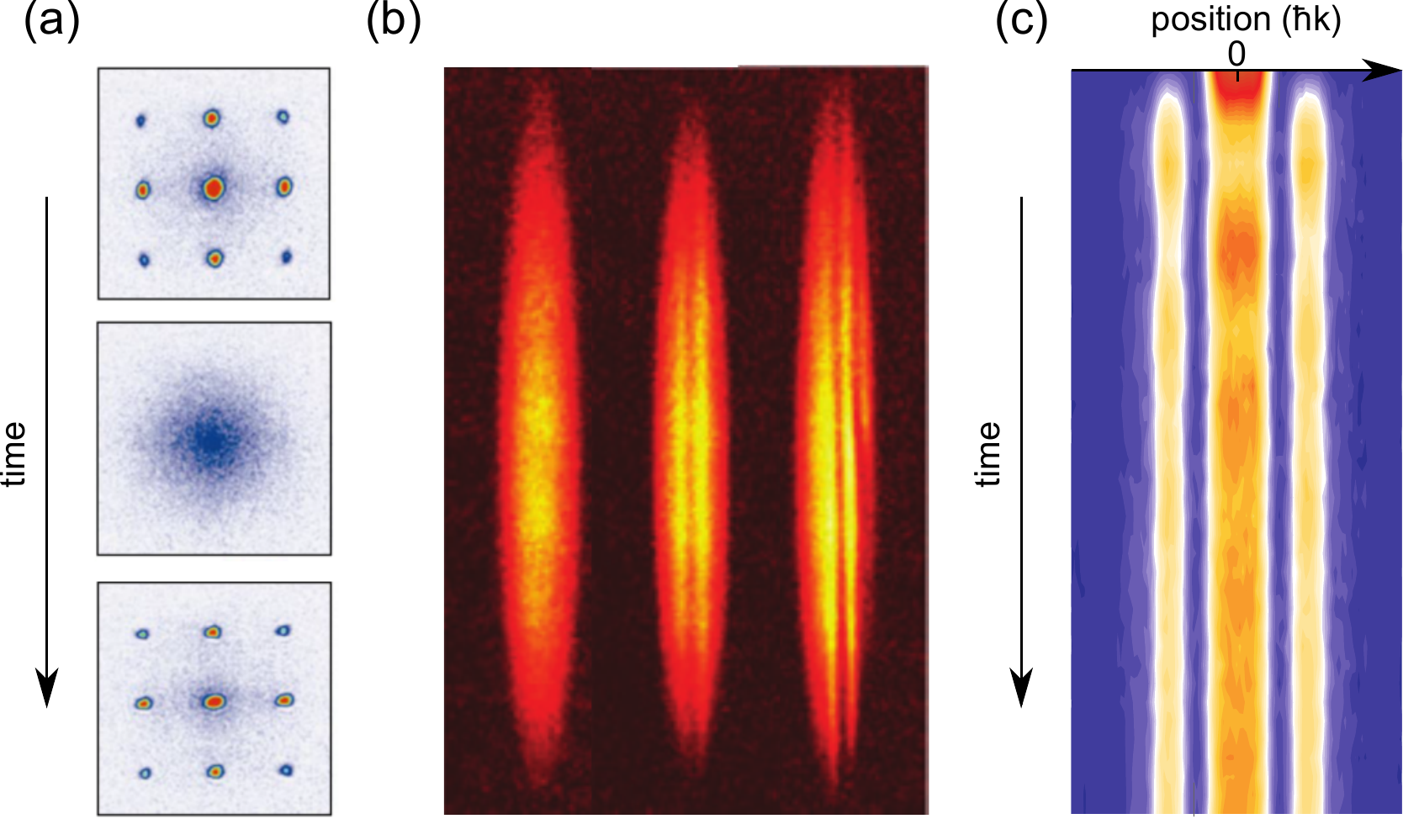}
	\caption{(a) Time-evolution of the momentum distribution of atoms in an optical lattice after a quench. Sharp peaks after time-of-flight expansion result from the coherent interference of atoms released from many different lattice sites. Collapses and revivals demonstrate the how the matterwave field periodically dephases and rephases during its unitary time evolution. Figure adapted from~\cite{Greiner02a}. (b) Time-of-flight images of atomic clouds after a quench across the threshold for Bose-Einstein condensation. From left to right, the clouds exhibit 0,1 and 2 defects. Figure adapted from~\cite{Lamporesi13}. (c) Dynamics of strongly-interacting bosons in an optical lattice. Zero (finite) momenta represent atoms on even (odd) lattice sites. Initializing the systen with all atoms on even sites, leads to an oscillatory relaxation towards a steady state with equally distributed atomic densities on even and odd sites. Figure adapted from~\cite{Trotzky12}.}
		\label{quench1}
\end{figure*}

The thermal equilibrium properties of atomic gases have been covered in a number of extensive reviews~\cite{PethickSmith,Leggett06,Castin01,BlochDalibardZwerger07}, both in the fermionic~\cite{KetterleFermi} and the bosonic~\cite{Ketterle99} case. Typically, realizations of such gases are very dilute with densities below $10^{14}$ atoms per $\text{cm}^{3}$ to avoid losses from molecule formation~\cite{Ketterle99}. These very low densities require ultracold temperatures to reach the quantum degenerate regime. These temperatures are typically achieved by a sequence of laser~\cite{Metcalf01} and evaporative cooling, with the gas being confined using magnetic or optical fields~\cite{Ketterle99}. The confinement can be tailored with great flexibility, from standard harmonic trapping potentials~\cite{Ketterle99,Grimm00} to micro-fabricated guides on atom chips~\cite{Folman00,Reichel11}, box-, ring- and micro-traps created using optical dipole potentials~\cite{Gaunt13b,Ramanathan11,Serwane11}, or optical lattices formed by interfering laser beams~\cite{Bloch05}. 

With typical system sizes from a few to $10^7$ atoms and temperatures reaching down to the picokelvin range, the mere existence of these gases demonstrates their near-perfect isolation from the environment. Moreover, only selected atomic species are introduced and cooled to ultracold temperatures~\cite{Schreck01,Taglieber08} and the atomic quantum states are under perfect control using optical, microwave or radio-frequency fields. At the same time, parameters like interaction strength, temperature, density, and dimensionality can be widely tuned, and the well-established and versatile techniques of atomic physics provide unique means for manipulation and probing~\cite{Ketterle99,PethickSmith}. In the following, we detail some of these techniques and give specific examples that provide insights into generic phenomena of non-equilibrium quantum many-body systems.  While many early experiments on ultracold atoms studied the rich and intricate collective dynamics of such gases that is not necessarily connected to relaxation processes and which can often be described using hydrodynamics and mean-field physics~\cite{Ketterle99,Moritz03,Bartenstein04}, we focus here on the connection between many-body dynamics and statistical physics.

\begin{figure*}[tb]
	\centering
		\includegraphics[width=.65\textwidth]{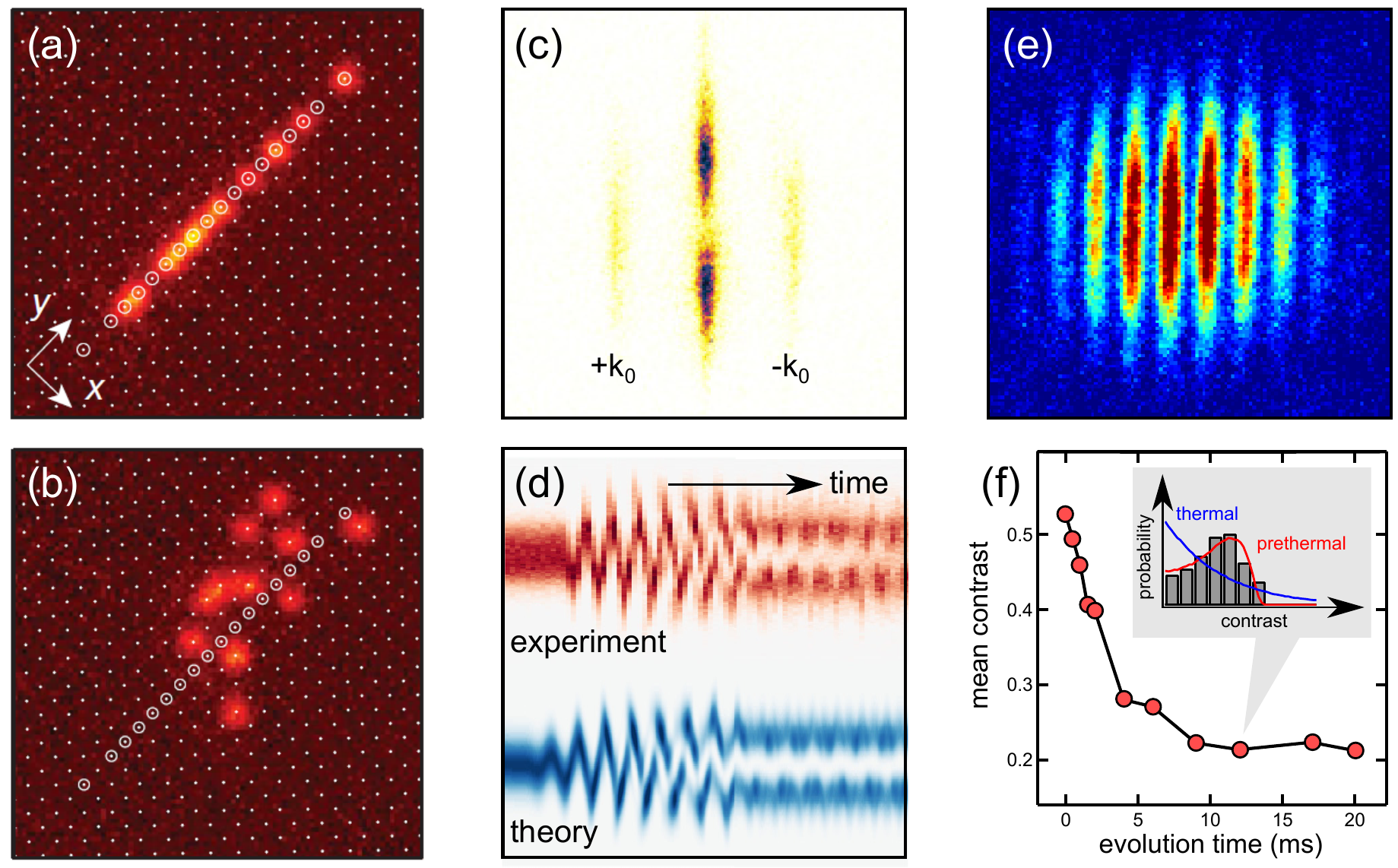}

	\caption{(a) \textit{In situ} fluorescence images of a line of single atoms prepared in an optical lattice. (b) The imaging technique enables the direct observation of the subsequent tunneling dynamics on the single atom level. Figures adapted from~\cite{Weitenberg2011}.  (c) Time-of-flight imaging of the momentum distribution, in this case also with single atom sensitivity using fluorescence imaging~\cite{Buecker09,Buecker11}. The high dynamic range of this technique enables the simultaneous imaging of both an high density initial cloud which has been prepared in the first excited state of the trap, and two much smaller clouds containing correlated atom pairs with momenta $\pm k_0$. (d) The creation and subsequent decay of the excited state are complex many-body processes. (e) Probing quantum fields using interference and absorption imaging in time-of-flight. (f) Studying the evolution of the mean contrast reveals the emergence of a steady states in the system's many-body dynamics, while the full counting statistics of the contrast enable the identification of this steady state as a prethermalized state~\cite{Gring12}.}
	\label{probing}
\end{figure*}

\subsection{Timescales}
The isolation from the environment and control over the quantum states of the atoms lead to very long coherence times that can reach many seconds. For studies of non-equilibrium properties, these coherence times have to be compared to the timescales of the dynamics. The latter are related to the relevant energy scales, like temperature, kinetic and interaction energy. They typically lie in the millisecond range, orders of magnitude below the coherence times. It is therefore possible to follow the intrinsic quantum dynamics of ultracold gases for very long times scales. From a more practical point, these dynamical timescales are also slow enough that no specialized ultrafast equipment is required in the experiments. 

In a landmark experiment Greiner et al.~\cite{Greiner02a,Will10} demonstrated the possibility of following the unitary dynamics of a quantum many-body system by studying atoms trapped in an optical lattice. In their experiment, the system was brought out of equilibrium by quenching the depth of the lattice potential within the superfluid regime. This created coherent superpositions of Fock states with different atom numbers on each lattice site. Because of the different energies associated with each Fock state the atomic matterwave field periodically dephased and rephased, resulting in long-lived coherent collapse and revival dynamics  (Fig.~\ref{quench1}a).

In a different series of non-equilibrium experiments, Weiler et al.~\cite{Weiler08} and Lamporesi et al.~\cite{Lamporesi13} monitored the actual formation of a Bose-Einstein condensate (BEC). In cold atom experiments, cooling across the BEC phase transition is typically achieved by evaporation, in which the most energetic atoms are removed from the gas. The remaining atoms subsequently rethermalize to a colder temperature through inter-atomic collisions. In the experiments, the removal of the particles was controlled using radio-frequency transitions to untrapped atomic states. This allowed to cross the phase transition with different velocities, revealing the formation of non-thermal states exhibiting defects (Fig.~\ref{quench1}b), in close analogy to the Kibble-Zurek mechanism~\cite{Zurek85}.

The coherence times in such experiments are in fact so long that their results can be challenging to reproduce using classical computers~\cite{Eisert10}. They are thus examples of dynamic quantum simulators~\cite{Feynman82,Bloch12,Nascimbene10,Trotzky12,Ku12,Vanhoucke12}, where well-controlled experimental dynamics is used to obtain results beyond the capabilities of present numerical quantum many-body simulations. An example is the experiment by Trotzky et al.~\cite{Trotzky12}, where the dynamics of a strongly-correlated state of bosonic atoms in an optical lattice was monitored (Fig.~\ref{quench1}c). Atomic densities, currents and coherences all showed a fast relaxation towards their equilibrium values and could be followed experimentally for times that were much longer than the ones accessible in a numerical time-dependent density-matrix renormalization group simulation~\cite{Schollwoeck05}. 

\subsection{Observables}
To probe their dynamics, ultracold gases can be imaged both in the confining trapping potential and after their release in time-of-flight expansion. While the former provides direct access to the density distribution $n(\bm r)$ in position space, the latter enables the measurement of the momentum distribution $n(\bm k)$. Typically, the detection process relies on the interaction of the atoms with laser light, either via absorption or fluorescence~\cite{Ketterle99,PethickSmith}. In terms of sensitivity, this detection process can reach the single atom level. For example, Sherson et al.~\cite{Sherson10} and Bakr et al.~\cite{Bakr10} prepared a Mott insulator~\cite{Greiner02} of bosonic ${}^{87}$Rb atoms trapped in the periodic potential of an optical lattice. Employing fluorescence imaging and high-resolution optics, they were able to image the system with both single lattice site and single atom resolution.

\begin{figure*}[tb]
	\centering
		\includegraphics[width=.7\textwidth]{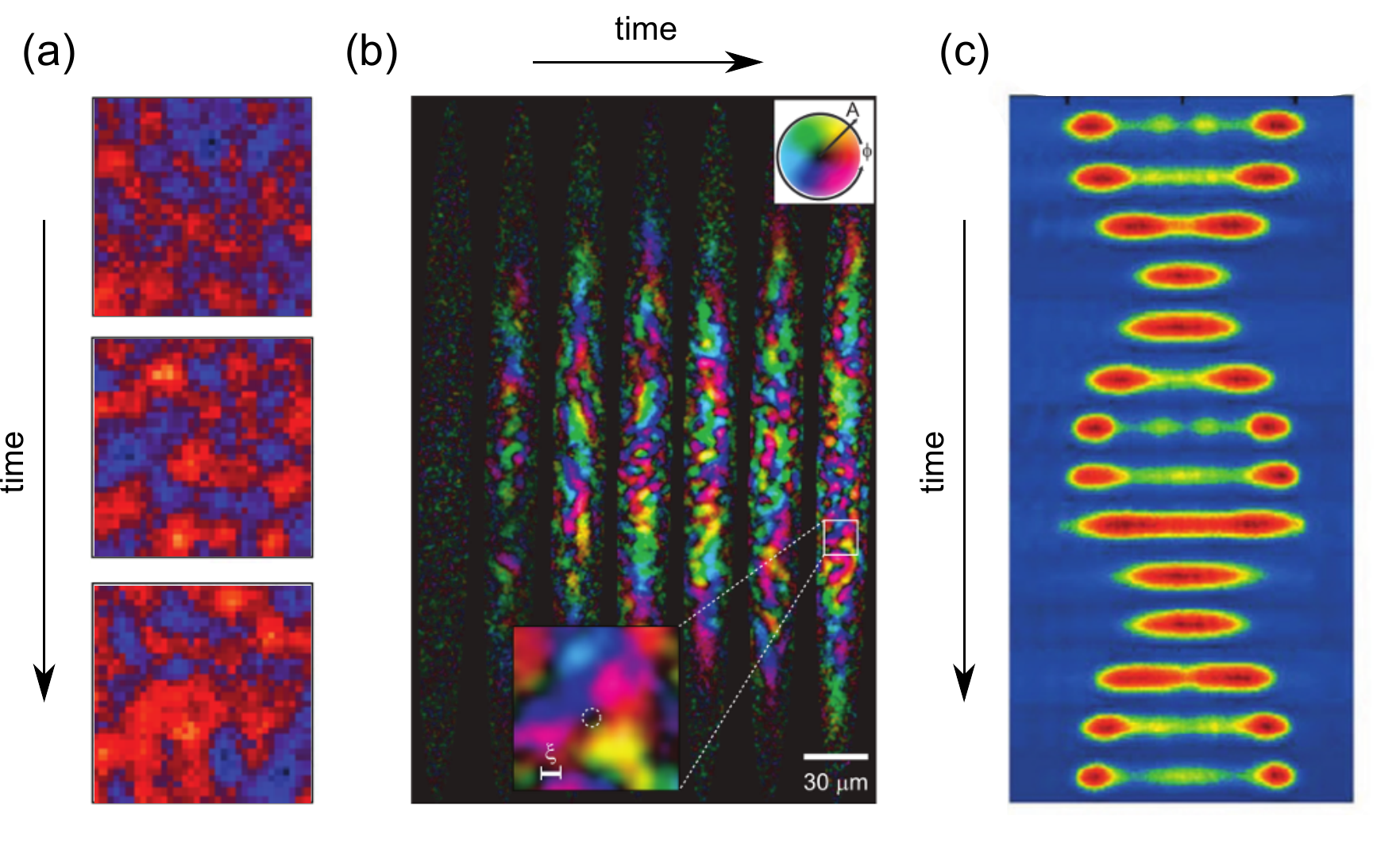}

	\caption{(a) Dynamics of density fluctuations after an interaction quench of a two-dimensional gas of cesium atoms. Figure adapted from~\cite{Hung13}. (b) Quench of a spinor gas, revealing the emergence of a steady state with long-lived spin structures. Figure adapted from~\cite{Sadler06} (c) Quantum Newton's cradle realized using 1D Bose gases. Long-lived oscillations in momentum space demonstrate the absence of thermalization in this integrable system. Figure adapted from~\cite{Kinoshita06}.}
	\label{fig:quench2}
\end{figure*}

Combining the flexibility of the lattice potential with this technique enabled a remarkable series of non-equilibrium experiments, for example a microscopic observation of atomic tunneling dynamics~\cite{Weitenberg2011} (see Fig.~\ref{probing}a,b), the spreading of quasi-particle pairs in a light-cone like dynamics~\cite{Cheneau12}, or the dynamics of two-magnon bound states~\cite{Fukuhara13}. 

Single atom sensitivity is not only limited to in-situ probes of ultracold atoms. Letting an expanding cloud of atoms pass through a resonant sheet of laser light, B\"ucker et al.~\cite{Buecker09} demonstrated single atom sensitive fluorescence imaging in time-of-flight. In the future, this scheme could be combined with focusing techniques~\cite{Jacqmin12} to yield not only single atom, but even single excitation sensitivity. 

Using this scheme enabled the observation of parametric amplification dynamics, leading to highly squeezed twin-atom beams~\cite{Buecker11} (for similar results, see also Refs.~\cite{Gross11,Luecke11}). By exploiting an optimal control pulse, atoms were transferred from the ground state of the trapping potential into the first excited state. This highly non-equilibrium state subsequently decayed back into the ground state via complex many-body dynamics which could be monitored with single atom sensitivity (Fig.~\ref{probing}c,d).

While the aforementioned imaging techniques can provide access to the density and momentum distributions, they do not allow the probing of the complex phase of the quantum fields under study. This can be accomplished using matterwave interference experiments where two ultracold gases are superimposed in time-of-flight expansion~\cite{Schumm05}. Such techniques are of high relevance for applications in metrology~\cite{Berrada13,Gross10,Riedel10}. 

For example, Gring et al.~\cite{Gring12} used the interference of two clouds of ${}^{87}$Rb atoms to show that this system retains memory of an initial non-equilibrium state for an extended time (Fig.~\ref{probing}e,f). A study of the full counting statistics of the interference contrast~\cite{Kitagawa10,Kuhnert13,Smith13,Langen13} revealed that instead of relaxing to thermal equilibrium, the system relaxed to a long-lived prethermalized state~\cite{Gring12,Kuhnert13,Berges04}. In this steady state, the system already showed thermal features like a temperature, but the state was still markedly different from thermal equilibrium.

Note that both fluorescence and absorption imaging are destructive methods. Experiments thus rely on the repeated preparation of many atomic clouds with identical initial conditions, which are let to evolve for different times. However, also less invasive techniques are possible, ranging from phase contrast imaging~\cite{Ketterle99}, imaging based on the Faraday effect~\cite{Gajdacz}, imaging of lattice gases using Raman sideband cooling~\cite{Patil14}, to more exotic techniques, such as electron microscopy of quantum gases~\cite{Gericke08b,Wuertz09}. One the one hand, these techniques enable a repeated probing of the same system. On the other hand, every measurement will inevitably influence the non-equilibrium dynamics through quantum back-action.

\subsection{Tunability of interactions}
One of the remarkable feature of ultracold atoms is the fact that interactions can be tuned over a wide range using Feshbach resonances~\cite{ChinRMP,Inouye98}. Such resonances arises in the scattering properties of the atoms when the state of two free atoms becomes resonant with a molecular bound state. Tuning the position of these two states with respect to each other can be achieved using magnetic fields. The interactions can be characterized using the s-wave scattering length $a$, which scales as $a(B) = a_{bg}[1-\Delta/(B-B_0)]$ across the resonance. Here, $B_0$ is the position of the resonance where $a$ diverges and changes sign, $\Delta$ is the width of the resonance, and $a_{bg}$ is the background scattering length which characterizes the scattering properties away from the resonance. Particularly convenient atomic species for such experiments are cesium~\cite{Vuletic99}, lithium~\cite{Junker08,Zuern13} and potassium~\cite{Loftus02,dErrico07}, where several wide Feshbach resonances are easily accessible in experiments. 

For example, Hung et al.~\cite{Hung13} started from a two-dimensional superfluid formed by loading a quantum degenerate sample of bosonic ${}^{133}$Cs atoms into a highly oblate trap. Subsequently, the system was taken out of equilibrium by quenching the interaction strength. The subsequent evolution of the non-equilibrium state showed two aspects. On long time scales, the system adjusted its overall density profile to the new interaction energy. On shorter time scales, density fluctuations in the cloud emerged, which were created by the sound waves generated in the interaction quench (Fig.~\ref{fig:quench2}a).

In another experiment Meinert et al.~\cite{Meinert13} used cesium atoms in a tilted optical lattice to implement a realization of the 1D Ising model with tunable interactions~\cite{Simon11,Sachdev02}. Changing the tilt of the lattice a sudden quench to the vicinity of the transition point of the Ising paramagnetic to anti-ferromagnetic quantum phase transition was realized. They observed coherent oscillations for the orientation of Ising spins, the properties of which could be widely tuned using a Feshbach resonance. Moreover this technique recently enabled the direct observation of higher-order tunneling processes over several lattice sites~\cite{Meinert14}.

Superheating was studied by Gaunt et al.~\cite{Gaunt13} using a gas of potassium atoms. This everyday non-equilibrium phenomenon occurs in many liquids such as water: when heated undisturbed the liquid does not boil, even above the boiling temperature. A very similar phenomenon can also occur in quantum systems, where a superheated Bose-Einstein condensate can persist above the critical temperature $T_c$. This superheated quantum state can be achieved by adjusting the atomic interactions of the potassium atoms so that thermalization in the BEC occurs faster than atoms are able to cross from the condensed to the thermal state, that is, to boil off from the BEC. In this remarkable state of a many-body quantum system, the condensed and uncondensed parts of the atom cloud have the same kinetic temperature, with the same average kinetic energy per atom. Nevertheless, the quantum degenerate BEC in the experiments persisted up to a temperature $50\%$ higher than $T_c$, and thermalization could be turned on or off by controlling the atomic scattering properties using a Feshbach resonance. Translated into our classical world this would correspond to observing superheated water at close to $575\,$K. The experiment thus showed that many-body quantum states can survive at higher temperatures for much longer than predicted by equilibrium physics.

\subsection{Internal degrees of freedom}

Atoms also contain well-controllable internal degrees of freedom which can be used to realize even richer physics. A prime example are spinor gases, where the atoms are prepared in hyperfine states with $F>0$. Here, $F$ denotes the quantum number of the total angular momentum. Such spinor gases can thus combine superfluid behavior and magnetism. They can also be used to study spin dynamics ~\cite{depaz13,Kronjaeger05,Widera08}. 

For the case of bosonic atoms with $F=1$ the system exhibits a quantum phase transition between a polar and a ferromagnetic phase. The competing parameters controlling this phase transition are the mean-field interaction of the spinor gas and a quadratic Zeeman interaction that can be tuned via an external magnetic field. Rapidly tuning an external magnetic field can be used to quench the system from one phase to the other. Such an experiment was realized by Sadler et al.~\cite{Sadler06}. During the subsequent time-evolution non-destructive phase contrast imaging was used to observe the dynamical emergence of magnetic domains in the gas. These were found to be very long-lived, indicating the emergence of a steady state. The observations are shown in (Fig.~\ref{fig:quench2}b).

The spin dynamics of spinor gases can also often be understood in analogy with a rigid pendulum. Such a pendulum features an unstable fixed point in its inverted position~\cite{Zibold10}. This is of particular interest as mean-field approximations fail in the vicinity of such fixed points because quantum fluctuations are strongly enhanced. Gerving et al.~\cite{Gerving12} realized such an inverted pendulum by initializing a gas of ${}^{87}$Rb atoms exactly at such an unstable fixed point and following its subsequent oscillatory dynamics. They then performed time-resolved measurements of the full counting statistics of the spin observables and found them to be in very good agreement with quantum calculations. Remarkably, for very long evolution times atom loss from their system increased the strength of the spin oscillations. This demonstrates how decoherence during the non-equilibrium dynamics of a many-body system can result in an increase of coherent behavior.

\subsection{Dimensionality, integrability and thermalization}
The confining trapping potentials for ultracold atoms can be made very strong such that the motion of the atoms is restricted to lower dimensions. In particular, one-dimensional (1D) Bose gases offer a model system which contains complex many-body physics but can still be captured with reasonable theoretical effort~\cite{Cazalilla11}. Moreover, the homogeneous 1D Bose gas with repulsive contact interactions is one of the hallmark examples of an integrable quantum system~\cite{Lieb63,Yang69}. The approximate realization of such a system in experiments thus allows the study of relaxation in the vicinity of multiple conserved quantities and hence the study of the interplay between integrability, many-body dynamics and thermalization. 

In cold atom experiments, a 1D Bose gas can be realized using anisotropic magnetic or optical trapping potentials, where the confinement in two directions is strong enough such that the temperature and the chemical potential of the system are smaller than the excited energy levels of the trapping potential~\cite{Krueger10,Esteve06}. This can be expressed by the condition $k_B T,\mu < \hbar \omega_\perp$, where $\omega_\perp$ denotes the harmonic trap frequency in the two strongly confining directions.

In contrast to the strongly confining directions many momentum modes can be occupied in the weakly confining direction. This leads to markedly different behavior than in 3D BECs, where only the lowest momentum mode is macroscopically occupied. These many momentum modes in 1D Bose gases are the origin of strong density and phase fluctuations, which prevent the creation of long-range order~\cite{Mermin66,Hohenberg67} and lead to a complex diagram of possible quantum states~\cite{Petrov00}.

A landmark non-equilibrium experiment based on such 1D Bose gases was realized by Kinoshita et al.~\cite{Kinoshita06}. They used atoms trapped in an optical lattice to realize an array of 1D Bose gases. The interaction parameter $\gamma=mg/\hbar n_\mathrm{1D}$ could be tuned over a wide range by changing the confining potential and the density $n_\mathrm{1D}$. Here, $m$ denotes the atomic mass and $g$ is the interaction strength, $\gamma \gg 1$ corresponds to a strongly-correlated Tonks-Girardeau gas~\cite{Kinoshita04,Paredes04}, $\gamma \ll 1$ to a weakly-interacting 1D Bose gas. By applying an optical phase grating~\cite{Cronin09}, a superposition of two momentum states with opposite sign was imposed on the gas. Given these initial conditions the atoms started to oscillate in momentum space, much like a Newton's cradle (see Fig.~\ref{fig:quench2}c). The resulting momentum distribution remained non-thermal even after thousands of collisions, for all realized interactions strengths. This was in stark contrast to the situation where the non-equilibrium momentum state was imposed on a 3D gas without the optical lattice. In the latter case, the system immediately relaxed to a thermal momentum distribution. The experiment thus confirmed that integrable or near-integrable systems need extremely long timescales to thermalize~\cite{Rigol09}. 

The integrability does not only affect the thermalization of the system, it also has strong effects on transport phenomena. This was observed by Ronzheimer et al.~\cite{Ronzheimer13,Sorg14}, studying the expansion of initially localized ultracold bosons in homogeneous one- and two-dimensional optical lattices. To this end ${}^{39}$K atoms were prepared in the combined potential of a 3D optical lattice and an additional harmonic confinement. Interactions could be tuned using a broad Feshbach resonance. The harmonic confinement was then decreased in one or two directions, so that the atoms could expand in a one- or two-dimensional optical lattice potential, realizing an interaction quench. It was observed that the fastest, ballistic expansion happened in all integrable limits of the system, where the presence of many constants of motion inhibited diffusive scattering. Deviations from these limits significantly suppressed the expansion and lead to signatures of diffusive dynamics. 

Finally, a series of experiments on the dynamics of 1D Bose gases was performed on atom chips~\cite{LangenPhD,Schaff14,Folman02,Reichel11}. In these experiments a single 1D Bose gas was studied, which enabled direct access to the fluctuation dynamics of the system. This situation is in contrast to the one in optical lattices, where the information contained in the fluctuations is inherently washed out by an average over many non-identical copies of the system. 

\begin{figure*}[tb]
	\centering
		\includegraphics[width=0.7\textwidth]{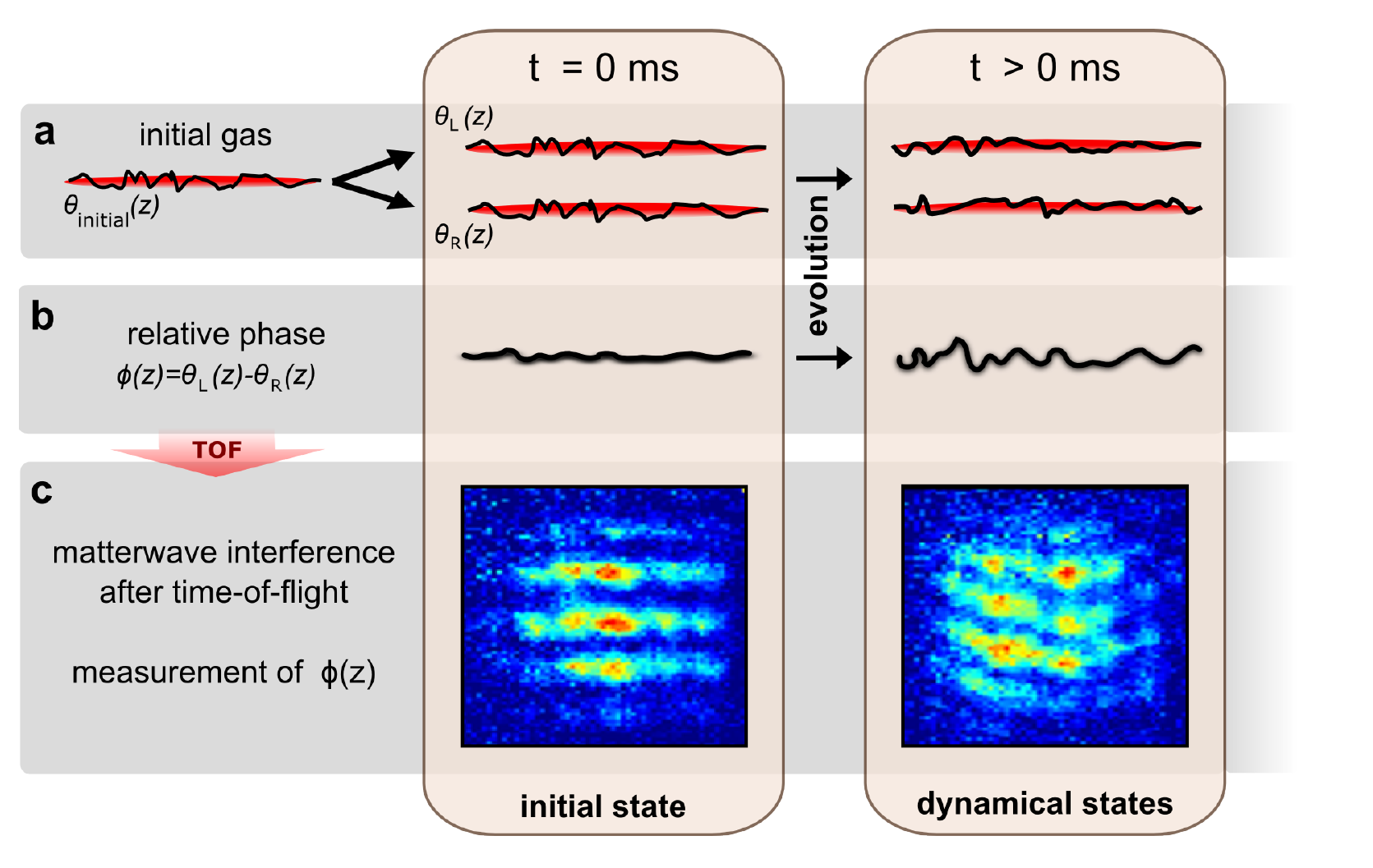}

	\caption{Experimental scheme for the study of non-equilibrium dynamics in a 1D Bose gas. (a) The experiment starts with a single gas in thermal equilibrium, where temperature manifests via phase fluctuations of the phase field $\theta_\mathrm{initial}(z)$ (represented by the black solid line).
This initial gas is quenched by splitting it into two uncoupled halves with almost identical phase fluctuations $\theta_L(z)$ and $\theta_R(z)$, and allowed to evolve for a time $t$ . (b) Consequently, at $t = 0\,$ms, fluctuations in the local phase difference $\phi(z)=\theta_L(z)-\theta_R(z)$ between the two gases are very small, but start to randomize for $t>0\,$ms. (c) shows typical experimental matter wave interference patterns obtained by overlapping the two gases in time-of-flight. Differences in the local relative phase lead to a locally displaced interference pattern, from which the relative phase, and thus the dynamics of the system, can be extracted by fitting a sinusoidal function to each pixel column. Figure adapted from~\cite{LangenPhD}.}
	\label{fig:expscheme}
\end{figure*}

Following up on the aforementioned observation of prethermalization~\cite{Gring12}, a quench was realized by transversely deforming the trapping potential into a fully tunable double well. This rapidly split the gas into two halves. The system was then let to evolve for a variable time. Subsequently, all trapping potentials were switched off, the gases expanded, overlapped and formed a matter wave interference pattern which could be imaged by standard techniques (see Fig.~\ref{fig:expscheme}). The interference pattern with its locally displaced interference fringes directly reflected the fluctuating relative phase field $\phi(z)$ between the two halves of the system. Immediately after the coherent splitting, the relative phase field was close to zero along the whole length of the system, resulting in straight fringes. Over time, the dynamics lead to a randomization of the relative phase and the corresponding interference patterns. This process can be quantified by studying the time evolution of two-point (or higher-order N-point phase correlation functions), given by $C(z_1,z_2) \sim \langle \Psi_1(z_1)\Psi_2^\dagger(z_1)\Psi_1^\dagger(z_2)\Psi_2(z_2)\rangle \approx \langle e^{i\phi(z_1)-i\phi(z_2)}\rangle$. Here, the $\Psi_{1,2}$ denote bosonic field operators describing the two halves of the system~\cite{Betz11,Langen13b}. The correlation functions contain only the measured relative phase $\phi(z)$ and can thus be directly calculated from the experimental data. 

\begin{figure}[tb]
	\centering
		\includegraphics[width=.43\textwidth]{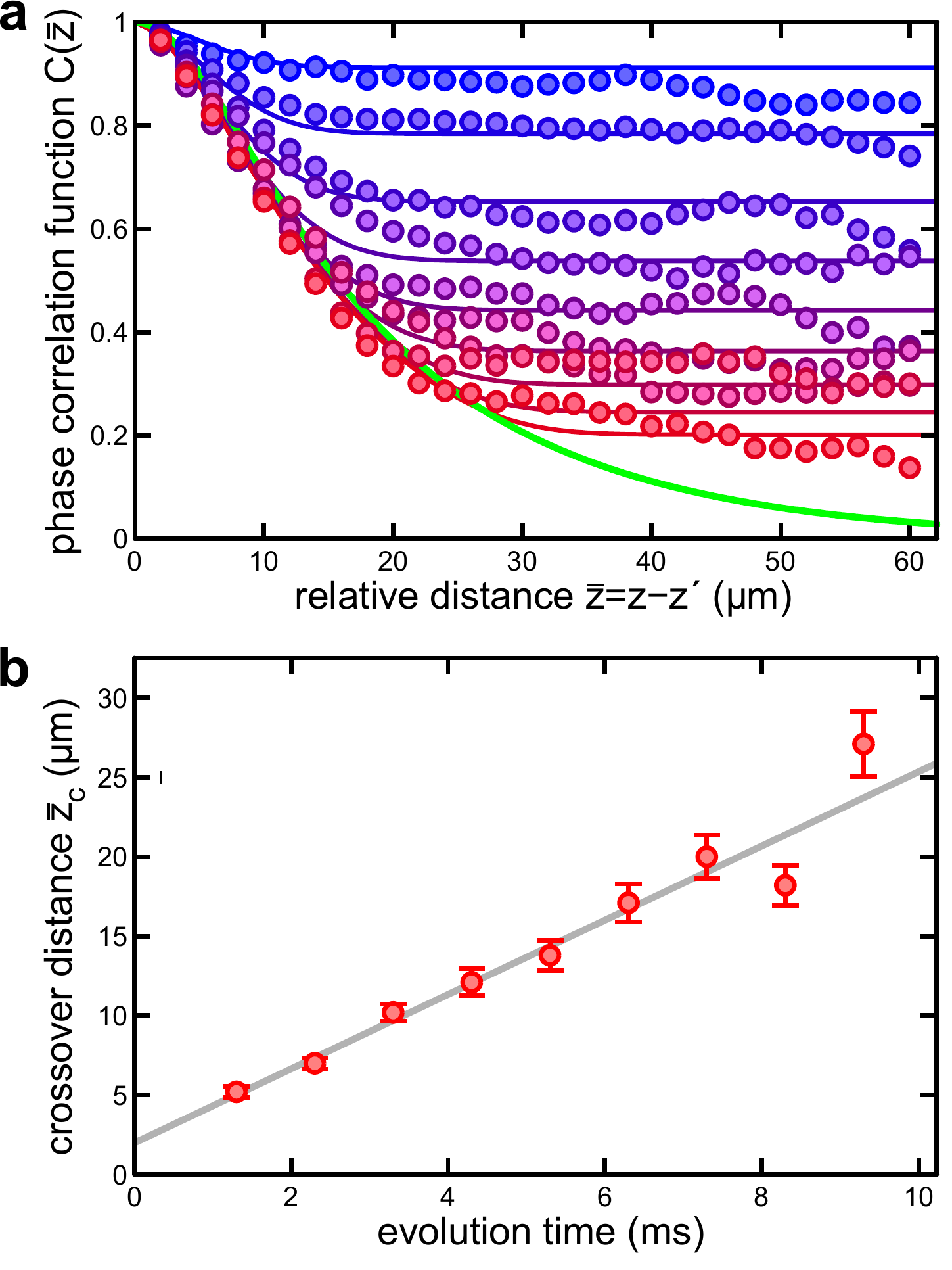}
	
		\caption{Local emergence of thermal correlations in a 1D Bose gas. \textbf{(a)} Experimental phase correlation functions $C(\bar z,t)$ (filled circles) compared to theoretical calculations (solid  lines). The evolution time $t$ increases from top to bottom, the bottom (green) line being the theoretical prediction for the relaxed, prethermalized state. For each $t$, correlation functions follow this prediction up to a crossover distance $\bar z_c(t)$ beyond which the system remembers the initial long-range phase coherence. \textbf{(b)} Position of the crossover distance $\bar z_c$ as a function of  $t$, revealing the light-cone-like emergence of the thermal correlations of the prethermalized state. The slope of the solid line corresponds to twice the phonon velocity of the system. Figure adapted from~\cite{Langen13b}.
		}
		\label{fig:lightcone}
\end{figure}

The dynamics of the two-point correlation function showed that the thermal correlations of a prethermalized state are first established locally and then spread through the system in a light-cone like evolution~\cite{Langen13b,Langen13,Geiger14} (Fig.~\ref{fig:lightcone}). In this context, the phononic excitations of the system could be interpreted as information carriers which propagate correlations through the system~\cite{Calabrese06,Cheneau12,Cramer08}. 

Moreover, using higher-order correlation functions the particular prethermalized state could be directly connected to a generalized Gibbs ensemble (GGE)~\cite{Langen14,Rigol07}. In its most general form such a GGE is described by the density matrix~\cite{Jaynes57b,Rigol07,Caux12} 
\begin{equation}
\hat \rho = \frac{1}{Z}\exp\left(-\sum_m\lambda_m\mathcal{\hat I}_m\right).\label{eq:gge}
\end{equation}
Here, $\{\mathcal{\hat I}_m\}$ denotes a full set of conserved quantities of the integrable system and $Z=\mathrm{Tr}[\exp(-\sum_m\lambda_m\mathcal{\hat I}_m)]$ is the partition function and $m\geq1$ is an integer. The numbers ${\lambda_m}$ are Lagrange multipliers associated with the conserved quantities. They are obtained by maximization of the entropy under the condition that the expectation values of the conserved quantities are fixed to their initial values. The GGE is a direct generalization of the usual thermodynamical ensembles and has been suggested to describe relaxed states in such diverse situations as integrable systems~\cite{Rigol07,Caux12}, localized systems~\cite{Vosk2013,Nandkishore14} or systems with approximately conserved quantities~\cite{Kollar11}. For example, in the case where only the energy is a conserved quantity, the GGE reduces to the standard canonical or Gibbs ensemble, with temperature being the only Lagrange multiplier. The GGE thus connects the relaxation of quantum many-body systems to the fundamental principle of entropy maximization, which forms the basis of statistical mechanics~\cite{Shannon:1949,Jaynes57,Jaynes57b}. From this, connections to the concepts of statistical physics that were outlined in the beginning can directly be established.

\section{Conclusion and perspectives}
The experiments with ultracold gases presented in this short review reveal a wide range of non-equilibrium phenomena. Yet, they are only a first step on the way to a general framework for non-equilibrium dynamics. This research is characterized by a close connection between experiment and theory. One particularly active line of research is the realization of textbook models of statistical physics to study their dynamics. Examples include Ising chains~\cite{Meinert13},
(super) Tonks-Girardeau~\cite{Paredes04,Kinoshita04,Haller09} and Yang-Yang gases~\cite{Amerongen08}, Luttinger liquids~\cite{Gring12,Haller10}, Hubbard- or Yang-Gaudin models~\cite{Stoeferle04,Greiner02,Bloch05,Moritz05,Jordens08,Schneider08}. Beyond the neutral atom gases discussed in this review, interesting systems have also been realized using charged ions~\cite{Ulm13,Jurcevic14,Richerme14}. 

Tunnel-coupled 1D bosons, as implemented using atom chips or optical lattices~\cite{Betz11,Greiner01b}, could be used to study the dynamics of the quantum Sine-Gordon model, which realizes a field theory with a gapped spectrum. In this case, interesting analogies with the dynamics of the early universe~\cite{Neuenhahn12a,Neuenhahn12,schmiedmayer13} and relativistic thermodynamics~\cite{Agarwal14} have been pointed out. Although it appears far-fetched to expect the full complexity and non-equilibrium dynamics of a high-energy or cosmological quantum field theory to be implemented on a degenerate quantum gas consisting only of a few thousand atoms, doing so may not be necessary. This relies on the important property encountered in quantum field theories that independence of the details of the underlying microscopic description can emerge when classical and quantum corrections are successively included across a range of scales~\cite{Wilson74}. This behavior exhibits universality which is well understood in thermal equilibrium. For example, at a continuous phase transition, characteristic scaling properties of microscopically very different systems can be characterized in terms of a few universality classes relating to common symmetry properties of the systems.
Similar considerations may apply also to non-equilibrium dynamics in many diverse areas of physics, ranging from the inflationary universe to complex quantum systems in condensed matter physics. Realistic proposals to observe such universal behavior away from thermal equilibrium using ultracold atoms have already been put forward~\cite{DallaTorre13,Novak12,Karl13}. We therefore expect future experiments to have profound implications for our understanding of the emergence of thermal and classical properties in isolated quantum many-body systems, the study of which is an ongoing theoretical and experimental endeavor. 
%%% Numbered Literature Cited

\section{Acknowledgements}
We acknowledge many stimulating discussions with our theoretical collegues Ehud Altman, J\"urgen Berges, Isabelle Bouchoule, Eugene Demler, Sebastian Diehl, Jens Eisert, Sebastian Erne, Thomas Gasenzer, Valentin Kasper, Takuya Kitagawa, Igor Mazets, Anatoli Polkovnikov, Marcos Rigol, as well as the members of our experimental group in Vienna. This work was supported by the EU (SIQS and ERC advanced grant QuantumRelax), the Austrian Science Fund (FWF) through the Sonderforschungsbereich FoQuS, the Doctoral Programme CoQuS (W1210) (T.L.) and the Lise Meitner Programme M1423 (R.G.).

\section{References}
\bibliographystyle{Science3}
\bibliography{biblio}

\begin{thebibliography}{100}

\bibitem{HuangBook}
K.~Huang, Statistical mechanics (Wiley, 1987).

\bibitem{Polkovnikov11}
A.~Polkovnikov, K.~Sengupta, A.~Silva, M.~Vengalattore, {\it Rev. Mod Phys\/}
  {\bf 83}, 863 (2011).

\bibitem{Podolsky06}
D.~Podolsky, G.~Felder, L.~Kofman, M.~Peloso, {\it Phys. Rev. D\/} {\bf 73},
  023501 (2006).

\bibitem{Kofman94}
L.~Kofman, A.~Linde, A.~A. Starobinsky, {\it Phys. Rev. Lett.\/} {\bf 73}, 3195
  (1994).

\bibitem{Arrizabalaga05}
A.~Arrizabalaga, J.~Smit, A.~Tranberg, {\it Phys. Rev. D\/} {\bf 72}, 025014
  (2005).

\bibitem{Berges04}
J.~Berges, S.~Bors\'anyi, C.~Wetterich, {\it Phys. Rev. Lett.\/} {\bf 93},
  142002 (2004).

\bibitem{BraunMunzinger01}
P.~Braun-Munzinger, D.~Magestro, K.~Redlich, S.~J., {\it Phys. Lett. B\/} {\bf
  518}, 41  (2001).

\bibitem{Heinz02}
U.~Heinz, P.~Kolb, {\it Nucl. Phys. A\/} {\bf 702}, 269 (2002).

\bibitem{Kollath07}
C.~Kollath, A.~M. L\"{a}uchli, E.~Altman, {\it Phys. Rev. Lett.\/} {\bf 98},
  180601 (2007).

\bibitem{Eckstein09}
M.~Eckstein, M.~Kollar, P.~Werner, {\it Phys. Rev. Lett.\/} {\bf 103}, 056403
  (2009).

\bibitem{Moeckel10}
M.~Moeckel, S.~Kehrein, {\it New J. Phys.\/} {\bf 12}, 055016 (2010).

\bibitem{Barnett10}
R.~Barnett, A.~Polkovnikov, M.~Vengalattore, {\it Phys. Rev. A\/} {\bf 84},
  023606 (2011).

\bibitem{Barmettler09}
P.~Barmettler, M.~Punk, V.~Gritsev, E.~Demler, E.~Altman, {\it Phys. Rev.
  Lett.\/} {\bf 102}, 130603 (2009).

\bibitem{Heyl13}
M.~Heyl, A.~Polkovnikov, S.~Kehrein, {\it Phys. Rev. Lett.\/} {\bf 110}, 135704
  (2013).

\bibitem{Muth10}
D.~Muth, B.~Schmidt, M.~Fleischhauer, {\it New J. Phys.\/} {\bf 12}, 083065
  (2010).

\bibitem{Srednicki94}
M.~Srednicki, {\it Phys. Rev. E\/} {\bf 50}, 888 (1994).

\bibitem{Rigol08}
M.~Rigol, V.~Dunjko, M.~Olshanii, {\it Nature\/} {\bf 452}, 854 (2008).

\bibitem{Reimann08}
P.~Reimann, {\it Phys. Rev. Lett.\/} {\bf 101}, 190403 (2008).

\bibitem{Popescu06}
S.~Popescu, A.~Short, A.~Winter, {\it Nature Phys.\/} {\bf 2}, 754 (2006).

\bibitem{PethickSmith}
C.~J. Pethick, H.~Smith, Bose-Einstein Condensation in Dilute Gases (Cambridge
  University Press, 2001).

\bibitem{Lamacraft12}
A.~Lamacraft, J.~Moore, {\it Ultracold Boson. Fermionic Gases\/} {\bf 5}, 177
  (2012).

\bibitem{LangenPhD}
T.~Langen, Non-equilibrium dynamics of one-dimensional bose gases, Ph.D.
  thesis, Vienna University of Technology (2012).

\bibitem{Neumann29}
J.~V. {Neumann}, {\it Z. Phys.\/} {\bf 57}, 30 (1929).

\bibitem{Deutsch91}
J.~M. Deutsch, {\it Phys. Rev. A\/} {\bf 43}, 2046 (1991).

\bibitem{Rigol12}
M.~Rigol, M.~Srednicki, {\it Phys. Rev. Lett.\/} {\bf 108}, 110601 (2012).

\bibitem{Cramer08}
M.~Cramer, C.~M. Dawson, J.~Eisert, T.~J. Osborne, {\it Phys. Rev. Lett.\/}
  {\bf 100}, 030602 (2008).

\bibitem{Rigol09}
M.~Rigol, {\it Phys. Rev. Lett.\/} {\bf 103}, 100403 (2009).

\bibitem{Nandkishore14}
R.~Nandkishore, D.~A. Huse, {\it arXiv:1404.0686\/}  (2014).

\bibitem{Kitagawa11}
T.~Kitagawa, A.~Imambekov, J.~Schmiedmayer, E.~Demler, {\it New. J. Phys.\/}
  {\bf 13}, 073018 (2011).

\bibitem{Kollar11}
M.~Kollar, F.~A. Wolf, M.~Eckstein, {\it Phys. Rev. B\/} {\bf 84}, 054304
  (2011).

\bibitem{Berges2007}
J.~Berges, T.~Gasenzer, {\it Phys. Rev. A\/} {\bf 76}, 033604 (2007).

\bibitem{Greiner02a}
M.~Greiner, O.~Mandel, T.~H\"{a}nsch, I.~Bloch, {\it Nature\/} {\bf 419}, 51
  (2002).

\bibitem{Lamporesi13}
G.~Lamporesi, S.~Donadello, S.~Serafini, F.~Dalfovo, G.~Ferrari, {\it Nature
  Physics\/} {\bf 9}, 656 (2013).

\bibitem{Trotzky12}
S.~Trotzky, {\it et~al.\/}, {\it Nature Phys.\/} {\bf 8}, 325 (2012).

\bibitem{Leggett06}
A.~J. Leggett, Quantum liquids: Bose condensation and Cooper pairing in
  condensed-matter systems (Oxford University Press, 2006).

\bibitem{Castin01}
Y.~Castin, {\it Lecture Notes of Les Houches Summer School\/}  (2001).

\bibitem{BlochDalibardZwerger07}
I.~Bloch, J.~Dalibard, W.~Zwerger, {\it Reviews of Modern Physics\/} {\bf 80},
  885 (2008).

\bibitem{KetterleFermi}
W.~Ketterle, M.~W. Zwierlein, {\it arXiv:0801.2500\/}  (2008).

\bibitem{Ketterle99}
W.~Ketterle, D.~Durfee, D.~Stamper-Kurn, {\it Bose-Einstein condensation in
  atomic gases, Proc. Internat. School Phys. Enrico Fermi\/}, M.~Inguscio,
  S.~Stringari, C.~Wieman, eds. (IOS Press, 1999), pp. 67--176.

\bibitem{Metcalf01}
H.~J. Metcalf, P.~van~der Straten, Laser Cooling and Trapping (Springer, 2001).

\bibitem{Grimm00}
R.~Grimm, M.~Weidem{\"u}ller, Y.~Ovchinnikov, {\it Adv. At. Mol. Opt. Phys.\/}
  {\bf 42}, 95 (2000).

\bibitem{Folman00}
R.~Folman, {\it et~al.\/}, {\it Phys. Rev. Lett.\/} {\bf 84}, 4749 (2000).

\bibitem{Reichel11}
J.~Reichel, V.~Vuletic, eds., Atom Chips (Wiley, VCH, 2011).

\bibitem{Gaunt13b}
A.~L. Gaunt, T.~F. Schmidutz, I.~Gotlibovych, R.~P. Smith, Z.~Hadzibabic, {\it
  Phys. Rev. Lett.\/} {\bf 110}, 200406 (2013).

\bibitem{Ramanathan11}
A.~Ramanathan, {\it et~al.\/}, {\it Phys. Rev. Lett.\/} {\bf 106}, 130401
  (2011).

\bibitem{Serwane11}
F.~{Serwane}, {\it et~al.\/}, {\it Science\/} {\bf 332}, 336 (2011).

\bibitem{Bloch05}
I.~Bloch, {\it Nature Physics\/} {\bf 1}, 23 (2005).

\bibitem{Schreck01}
F.~Schreck, {\it et~al.\/}, {\it Physical Review Letters\/} {\bf 87}, 080403
  (2001).

\bibitem{Taglieber08}
M.~Taglieber, A.-C. Voigt, T.~Aoki, T.~H{\"a}nsch, K.~Dieckmann, {\it Physical
  review letters\/} {\bf 100}, 010401 (2008).

\bibitem{Moritz03}
H.~Moritz, T.~St\"oferle, M.~K\"ohl, T.~Esslinger, {\it Phys. Rev. Lett.\/}
  {\bf 91}, 250402 (2003).

\bibitem{Bartenstein04}
M.~Bartenstein, {\it et~al.\/}, {\it Phys. Rev. Lett.\/} {\bf 92}, 203201
  (2004).

\bibitem{Weitenberg2011}
C.~Weitenberg, {\it et~al.\/}, {\it Nature\/} {\bf 471}, 319 (2011).

\bibitem{Buecker09}
R.~B\"{u}cker, {\it et~al.\/}, {\it New J. Phys.\/} {\bf 11}, 103039 (2009).

\bibitem{Buecker11}
R.~B{\"u}cker, {\it et~al.\/}, {\it Nature Physics\/} {\bf 7}, 608 (2011).

\bibitem{Gring12}
M.~Gring, {\it et~al.\/}, {\it Science\/} {\bf 337}, 1318 (2012).

\bibitem{Will10}
S.~Will, {\it et~al.\/}, {\it Nature\/} {\bf 465}, 197 (2010).

\bibitem{Weiler08}
C.~N. Weiler, {\it et~al.\/}, {\it Nature\/} {\bf 455}, 948 (2008).

\bibitem{Zurek85}
W.~Zurek, {\it Nature\/} {\bf 317}, 505 (1985).

\bibitem{Eisert10}
J.~Eisert, M.~Cramer, M.~B. Plenio, {\it Reviews of Modern Physics\/} {\bf 82},
  277 (2010).

\bibitem{Feynman82}
R.~P. Feynman, {\it International journal of theoretical physics\/} {\bf 21},
  467 (1982).

\bibitem{Bloch12}
I.~Bloch, J.~Dalibard, S.~Nascimb{\`e}ne, {\it Nature Physics\/} {\bf 8}, 267
  (2012).

\bibitem{Nascimbene10}
S.~Nascimb\`ene, N.~Navon, K.~Jiang, F.~Chevy, C.~Salomon, {\it Nature\/} {\bf
  463}, 1057 (2010).

\bibitem{Ku12}
M.~J.~H. Ku, A.~T. Sommer, L.~W. Cheuk, M.~W. Zwierlein, {\it Science\/} {\bf
  335}, 563 (2012).

\bibitem{Vanhoucke12}
K.~Van~Houcke, {\it et~al.\/}, {\it Nature Phys.\/} {\bf 8}, 366 (2012).

\bibitem{Schollwoeck05}
U.~Schollw{\"o}ck, {\it Reviews of Modern Physics\/} {\bf 77}, 259 (2005).

\bibitem{Sherson10}
J.~F. {Sherson}, {\it et~al.\/}, {\it Nature\/} {\bf 467}, 68 (2010).

\bibitem{Bakr10}
W.~S. {Bakr}, {\it et~al.\/}, {\it Science\/} {\bf 329}, 547 (2010).

\bibitem{Greiner02}
M.~Greiner, O.~Mandel, T.~Esslinger, T.~W. H\"{a}nsch, I.~Bloch, {\it Nature\/}
  {\bf 415}, 39 (2002).

\bibitem{Hung13}
C.-L. Hung, V.~Gurarie, C.~Chin, {\it Science\/} {\bf 341}, 1213 (2013).

\bibitem{Sadler06}
L.~E. Sadler, J.~M. Higbie, S.~R. Leslie, M.~Vengalattore, D.~M. Stamper-Kurn,
  {\it Nature\/} {\bf 443}, 312 (2006).

\bibitem{Kinoshita06}
T.~Kinoshita, T.~Wenger, D.~Weiss, {\it Nature\/} {\bf 440}, 900 (2006).

\bibitem{Cheneau12}
M.~Cheneau, {\it et~al.\/}, {\it Nature\/} {\bf 481}, 484 (2012).

\bibitem{Fukuhara13}
T.~{Fukuhara}, {\it et~al.\/}, {\it Nature\/} {\bf 502}, 76 (2013).

\bibitem{Jacqmin12}
T.~Jacqmin, B.~Fang, T.~Berrada, T.~Roscilde, I.~Bouchoule, {\it Phys. Rev.
  A\/} {\bf 86}, 043626 (2012).

\bibitem{Gross11}
C.~Gross, {\it et~al.\/}, {\it Nature\/} {\bf 480}, 219 (2011).

\bibitem{Luecke11}
B.~L{\"u}cke, {\it et~al.\/}, {\it Science\/} {\bf 334}, 773 (2011).

\bibitem{Schumm05}
T.~Schumm, {\it et~al.\/}, {\it Nature Phys.\/} {\bf 1}, 57 (2005).

\bibitem{Berrada13}
T.~Berrada, {\it et~al.\/}, {\it Nat. Comm.\/} p. 2077 (2013).

\bibitem{Gross10}
C.~Gross, T.~Zibold, E.~Nicklas, J.~Esteve, M.~K. Oberthaler, {\it Nature\/}
  {\bf 464}, 1165 (2010).

\bibitem{Riedel10}
M.~F. Riedel, {\it et~al.\/}, {\it Nature\/} {\bf 464}, 1170 (2010).

\bibitem{Kitagawa10}
T.~Kitagawa, {\it et~al.\/}, {\it Phys. Rev. Lett.\/} {\bf 104}, 255302 (2010).

\bibitem{Kuhnert13}
M.~Kuhnert, {\it et~al.\/}, {\it Phys. Rev. Lett.\/} {\bf 110}, 090405 (2013).

\bibitem{Smith13}
D.~{Adu Smith}, {\it et~al.\/}, {\it New J. Phys.\/} {\bf 15}, 075011 (2013).

\bibitem{Langen13}
T.~Langen, {\it et~al.\/}, {\it Eur. Phys. J. Special Topics\/} {\bf 217}, 43
  (2013).

\bibitem{Gajdacz}
M.~Gajdacz, {\it et~al.\/}, {\it Review of Scientific Instruments\/} {\bf 84},
  (2013).

\bibitem{Patil14}
Y.~Patil, L.~Aycock, S.~Chakram, M.~Vengalattore, {\it arXiv:1404.5583\/}
  (2014).

\bibitem{Gericke08b}
T.~Gericke, P.~W{\"u}rtz, D.~Reitz, T.~Langen, H.~Ott, {\it Nature Physics\/}
  {\bf 4}, 949 (2008).

\bibitem{Wuertz09}
P.~W\"urtz, T.~Langen, T.~Gericke, A.~Koglbauer, H.~Ott, {\it Phys. Rev.
  Lett.\/} {\bf 103}, 080404 (2009).

\bibitem{ChinRMP}
C.~Chin, R.~Grimm, P.~Julienne, E.~Tiesinga, {\it Rev. Mod. Phys.\/} {\bf 82},
  1225 (2010).

\bibitem{Inouye98}
S.~Inouye, {\it et~al.\/}, {\it Nature\/} {\bf 392}, 151 (1998).

\bibitem{Vuletic99}
V.~Vuleti{\'c}, A.~J. Kerman, C.~Chin, S.~Chu, {\it Physical review letters\/}
  {\bf 82}, 1406 (1999).

\bibitem{Junker08}
M.~Junker, {\it et~al.\/}, {\it Phys. Rev. Lett.\/} {\bf 101}, 060406 (2008).

\bibitem{Zuern13}
G.~Z\"urn, {\it et~al.\/}, {\it Phys. Rev. Lett.\/} {\bf 110}, 135301 (2013).

\bibitem{Loftus02}
T.~Loftus, C.~A. Regal, C.~Ticknor, J.~L. Bohn, D.~S. Jin, {\it Phys. Rev.
  Lett.\/} {\bf 88}, 173201 (2002).

\bibitem{dErrico07}
C.~D'Errico, {\it et~al.\/}, {\it New Journal of Physics\/} {\bf 9}, 223
  (2007).

\bibitem{Meinert13}
F.~Meinert, {\it et~al.\/}, {\it Phys. Rev. Lett.\/} {\bf 111}, 053003 (2013).

\bibitem{Simon11}
J.~{Simon}, {\it et~al.\/}, {\it Nature\/} {\bf 472}, 307 (2011).

\bibitem{Sachdev02}
S.~Sachdev, K.~Sengupta, S.~M. Girvin, {\it Phys. Rev. B\/} {\bf 66}, 075128
  (2002).

\bibitem{Meinert14}
F.~Meinert, {\it et~al.\/}, {\it Science\/} {\bf 344}, 1259 (2014).

\bibitem{Gaunt13}
A.~L. Gaunt, R.~J. Fletcher, R.~P. Smith, Z.~Hadzibabic, {\it Nature Phys\/}
  {\bf 9}, 271 (2013).

\bibitem{depaz13}
A.~de~Paz, {\it et~al.\/}, {\it Physical review letters\/} {\bf 111}, 185305
  (2013).

\bibitem{Kronjaeger05}
J.~{Kronj{\"a}ger}, {\it et~al.\/}, {\it Phys. Rev. A\/} {\bf 72}, 063619
  (2005).

\bibitem{Widera08}
A.~Widera, {\it et~al.\/}, {\it Phys. Rev. Lett.\/} {\bf 100}, 140401 (2008).

\bibitem{Zibold10}
T.~Zibold, E.~Nicklas, C.~Gross, M.~K. Oberthaler, {\it Physical review
  letters\/} {\bf 105}, 204101 (2010).

\bibitem{Gerving12}
C.~S. Gerving, {\it et~al.\/}, {\it Nature Communications\/} {\bf 3}, 1169
  (2012).

\bibitem{Cazalilla11}
M.~A. Cazalilla, R.~Citro, T.~Giamarchi, E.~Orignac, M.~Rigol, {\it Rev. Mod.
  Phys.\/} {\bf 83}, 1405 (2011).

\bibitem{Lieb63}
E.~H. Lieb, W.~Liniger, {\it Phys. Rev.\/} {\bf 130}, 1605 (1963).

\bibitem{Yang69}
C.~N. Yang, {\it J. Mat. Phys.\/} {\bf 10}, 1115 (1969).

\bibitem{Krueger10}
P.~Kr\"uger, S.~Hofferberth, I.~E. Mazets, I.~Lesanovsky, J.~Schmiedmayer, {\it
  Phys. Rev. Lett.\/} {\bf 105}, 265302 (2010).

\bibitem{Esteve06}
J.~Esteve, {\it et~al.\/}, {\it Phys. Rev. Lett.\/} {\bf 96}, 130403 (2006).

\bibitem{Mermin66}
N.~D. Mermin, H.~Wagner, {\it Phys. Rev. Lett.\/} {\bf 17}, 1133 (1966).

\bibitem{Hohenberg67}
P.~C. Hohenberg, {\it Phys. Rev.\/} {\bf 158}, 383 (1967).

\bibitem{Petrov00}
D.~S. Petrov, G.~V. Shlyapnikov, J.~T.~M. Walraven, {\it Phys. Rev. Lett.\/}
  {\bf 85}, 3745 (2000).

\bibitem{Kinoshita04}
T.~{Kinoshita}, T.~{Wenger}, D.~S. {Weiss}, {\it Science\/} {\bf 305}, 1125
  (2004).

\bibitem{Paredes04}
B.~{Paredes}, {\it et~al.\/}, {\it Nature\/} {\bf 429}, 277 (2004).

\bibitem{Cronin09}
A.~Cronin, J.~Schmiedmayer, D.~Pritchard, {\it Rev. Mod. Phys.\/} {\bf 81},
  1051 (2009).

\bibitem{Ronzheimer13}
J.~P. Ronzheimer, {\it et~al.\/}, {\it Phys. Rev. Lett.\/} {\bf 110}, 205301
  (2013).

\bibitem{Sorg14}
S.~Sorg, L.~Vidmar, L.~Pollet, F.~Heidrich-Meisner, {\it arXiv:1405.5404\/}
  (2014).

\bibitem{Schaff14}
J.-F. Schaff, T.~Langen, J.~Schmiedmayer. {to appear in Lecture Notes of the
  Varenna Summer school 2013.}

\bibitem{Folman02}
R.~Folman, P.~Kruger, J.~Schmiedmayer, J.~Denschlag, C.~Henkel, {\it Adv. At.
  Mol. Opt. Phys.\/} {\bf 48}, 263 (2002).

\bibitem{Betz11}
T.~Betz, {\it et~al.\/}, {\it Phys. Rev. Lett.\/} {\bf 106}, 020407 (2011).

\bibitem{Langen13b}
T.~{Langen}, R.~Geiger, M.~{Kuhnert}, B.~{Rauer}, J.~Schmiedmayer, {\it Nature
  Phys.\/} {\bf 9}, 640 (2013).

\bibitem{Geiger14}
R.~Geiger, T.~Langen, I.~Mazets, J.~Schmiedmayer, {\it New Journal of
  Physics\/} {\bf 16}, 053034 (2014).

\bibitem{Calabrese06}
P.~Calabrese, J.~Cardy, {\it Phys. Rev. Lett.\/} {\bf 96}, 011368 (2006).

\bibitem{Langen14}
T.~L. et~al., {\it submitted\/}  (2014).

\bibitem{Rigol07}
M.~Rigol, V.~Dunjko, V.~Yurovsky, M.~Olshanii, {\it Phys. Rev. Lett.\/} {\bf
  98}, 050405 (2007).

\bibitem{Jaynes57b}
E.~T. Jaynes, {\it Phys. Rev.\/} {\bf 108}, 171 (1957).

\bibitem{Caux12}
J.-S. Caux, R.~M. Konik, {\it Phys. Rev. Lett.\/} {\bf 109}, 175301 (2012).

\bibitem{Vosk2013}
R.~Vosk, E.~Altman, {\it Phys. Rev. Lett.\/} {\bf 110}, 067204 (2013).

\bibitem{Shannon:1949}
C.~E. Shannon, W.~Weaver, The mathematical theory of communication (The
  University of Illinois Press, Urbana, IL, 1949).

\bibitem{Jaynes57}
E.~T. Jaynes, {\it Phys. Rev.\/} {\bf 106}, 620 (1957).

\bibitem{Haller09}
E.~Haller, {\it et~al.\/}, {\it Science\/} {\bf 325}, 1224 (2009).

\bibitem{Amerongen08}
A.~H. van Amerongen, J.~J.~P. van Es, P.~Wicke, K.~V. Kheruntsyan, N.~J. van
  Druten, {\it Phys. Rev. Lett.\/} {\bf 100}, 090402 (2008).

\bibitem{Haller10}
E.~Haller, {\it et~al.\/}, {\it Nature\/} {\bf 466}, 597 (2010).

\bibitem{Stoeferle04}
T.~{St{\"o}ferle}, H.~{Moritz}, C.~{Schori}, M.~{K{\"o}hl}, T.~{Esslinger},
  {\it Phys. Rev. Lett.\/} {\bf 92}, 130403 (2004).

\bibitem{Moritz05}
H.~Moritz, T.~St\"oferle, K.~G\"unter, M.~K\"ohl, T.~Esslinger, {\it Phys. Rev.
  Lett.\/} {\bf 94}, 210401 (2005).

\bibitem{Jordens08}
R.~J{\"o}rdens, N.~Strohmaier, K.~G{\"u}nter, H.~Moritz, T.~Esslinger, {\it
  Nature\/} {\bf 455}, 204 (2008).

\bibitem{Schneider08}
U.~Schneider, {\it et~al.\/}, {\it Science\/} {\bf 322}, 1520 (2008).

\bibitem{Ulm13}
S.~Ulm, {\it et~al.\/}, {\it Nature communications\/} {\bf 4} (2013).

\bibitem{Jurcevic14}
P.~Jurcevic, {\it et~al.\/}, {\it Nature (in press)\/}  (2014).

\bibitem{Richerme14}
P.~Richerme, {\it et~al.\/}, {\it Nature (in press), arXiv:1401.5088\/}
  (2014).

\bibitem{Greiner01b}
M.~Greiner, I.~Bloch, O.~Mandel, T.~W. H{\"a}nsch, T.~Esslinger, {\it Phys.
  Rev. Lett.\/} {\bf 8716}, 160405 (2001).

\bibitem{Neuenhahn12a}
C.~Neuenhahn, A.~Polkovnikov, F.~Marquardt, {\it Physical review letters\/}
  {\bf 109}, 085304 (2012).

\bibitem{Neuenhahn12}
C.~Neuenhahn, F.~Marquardt, {\it arXiv:1208.2255\/}  (2012).

\bibitem{schmiedmayer13}
J.~Schmiedmayer, J.~Berges, {\it Science\/} {\bf 341}, 1188 (2013).

\bibitem{Agarwal14}
K.~Agarwal, {\it et~al.\/}, {\it arXiv:1402.6716\/}  (2014).

\bibitem{Wilson74}
K.~G. Wilson, J.~Kogut, {\it Physics Reports\/} {\bf 12}, 75 (1974).

\bibitem{DallaTorre13}
E.~G. Dalla~Torre, E.~Demler, A.~Polkovnikov, {\it Phys. Rev. Lett.\/} {\bf
  110}, 090404 (2013).

\bibitem{Novak12}
B.~Nowak, J.~Schole, D.~Sexty, T.~Gasenzer, {\it Physical Review A\/} {\bf 85},
  043627 (2012).

\bibitem{Karl13}
M.~Karl, B.~Nowak, T.~Gasenzer, {\it Scientific reports\/} {\bf 3} (2013).

\end{thebibliography}

\end{document}